%% file: OME234_LIF_LII_mech.tex
\documentclass[11pt,a4paper,onecolumn,reqno]{amsart}
\usepackage[a4paper, margin=2.3cm, bmargin=3cm]{geometry}

\usepackage{graphicx,stfloats}
\usepackage{epstopdf}
\usepackage{color}
\usepackage{amsmath}
\usepackage{amsaddr}
\usepackage{bm, soul}
\usepackage{mathtools}
\usepackage[colorinlistoftodos,prependcaption,textsize=tiny]{todonotes}
\usepackage{braket}
\usepackage{hyperref}
\usepackage{makecell}
\usepackage{array}
\usepackage{placeins}
\usepackage[square,comma,sort&compress, numbers]{natbib}

\usepackage[version=3]{mhchem} \usepackage{siunitx}
 \usepackage{multirow}

\usepackage{etoolbox}
\patchcmd{\pprintMaketitle}
  {\hrule\vskip12pt}
 {\hrule\vskip12pt\ifvoid\extrainfobox\else\unvbox\extrainfobox\par\vskip12pt\fi}
 {}{}

\newsavebox\extrainfobox

\usepackage{caption}
\title{Particle formation in oxymethylene ethers \\ (\ce{OME_n}, n = 2-4) / ethylene premixed flames}
\author[stfs]{Robert Schmitz$^1$, Federica Ferraro$^{1,*}$, Mariano Sirignano$^2$, Christian Hasse$^1$ }
\email{ferraro@stfs.tu-darmstadt.de} 
\address[]{$^1$ Technical University of Darmstadt, Department of Mechanical Engineering, Simulation of reactive Thermo-Fluid Systems, Otto-Berndt-Str. 2, 64287 Darmstadt, Germany \\
$^2$Dipartimento di Ingegneria Chimica, dei Materiali e della Produzione Industriale – Universita' degli Studi di Napoli Federico II, P. le Tecchio 80, 80125 Napoli, Italy}

\begin{document}
\pagestyle{plain}
\maketitle


\begin{abstract} 
Alternative synthetic fuels can be produced by renewable energy sources and represent a potential route for solving long-term energy storage.
Among them, oxygenated fuels have the advantage of significantly reducing pollutant emissions and can therefore  be used as carbon-neutral substitute fuels for transportation.
In this work, the sooting propensity of different oxymethylene ethers (\ce{OME_n}) is investigated using a combined experimental and numerical study on a series of burner-stabilized premixed flames under mild to severe sooting conditions. 
Herein, mixtures of ethylene in combination with the three individual \ce{OME_n} for n $=$2-4 are compared in terms of  soot formation behavior with  pure ethylene flames. 
The kinetic mechanism from Sun et al. (Proc. Combust. Inst. 36, 1269-1278, 2017)  for oxymethylene ether \ce{OME_n} combustion with n$=${1-3} is extended to include \ce{OME4} decomposition and combustion kinetics.
In the numerical simulations, the kinetic mechanism is combined with a detailed quadvariate soot model which uses the Conditional Quadrature Method of Moments and validated with Laser-Induced Fluorescence (LIF) and Laser-Induced Incandescence (LII) measurements.
It is observed, that the three investigated \ce{OME_n} with n$=${2,3,4} show similar sooting behavior, mainly reducing larger aggregates while not significantly affecting the formation of smaller particles. Furthermore, the extent of soot reduction is comparable among the three \ce{OME_n}. The trends and overall reduction are very well captured by the model.
The modeling results are analyzed through reaction path analyses and sensitivity studies which show the important role of \ce{OME_n} decomposition  and the formation of \ce{CH2O} under these rich conditions to reduce species relevant for soot formation.
This is at the  base of the negligible observed differences in terms of soot formation for the different \ce{OME_n} fuels.
\end{abstract}



\clearpage

\section{Introduction} \label{S:I} \addvspace{10pt}
\input{parts/introduction}

\section{Experimental Setup} \label{S:E} \addvspace{10pt}
\input{parts/exp_setup}

\section{Numerical Modeling} \label{S:N} \addvspace{10pt}
\input{parts/num_setup}

\section{Results}  \label{S:R} \addvspace{10pt}
\input{parts/results}

\section{Conclusions} \label{S:C} \addvspace{10pt}
\input{parts/conclusions}

\section*{Acknowledgments} \label{Acknowledgments}
The authors gratefully acknowledge the funding by the German Federal Ministry of Education and Research (BMBF) as part of the NAMOSYN Project (project number 03SF0566R0). 


\bibliography{library.bib}
\bibliographystyle{unsrtnat}
\end{document}

%% file: parts/introduction.tex
Oxygenated synthetic fuels  can support the decarbonization of practical combustion devices reducing pollutant emissions especially in hard to electrify sectors such as maritime, heavy-duty and air transportation. 
Oxymethylene ethers (OMEs), \ce{CH3O(CH2O)_nCH3},  also known as polyoxymethylene dimethyl ethers (PODEs or DMMs), are promising fuel candidates and are currently being investigated for self-ignition engine applications. They can be produced by renewable energy sources~\cite{Fenard2021} yielding a neutral greenhouse gas balance. Recent studies have shown that OME fuels, pure or in blends, can significantly reduce carbon monoxide (\ce{CO}), unburned hydrocarbons, and soot particles, e.g.,~\cite{Wang2016, Liu2016,Ferraro2021,Schmitz2021,Tan2021}. Their molecular structure exhibits no C-C bonds, resulting in a fast  oxidation process, a reduced amount of gas-phase particle precursors and particle formation. 

Although  combustion of pure OMEs yields an almost complete soot suppression~\cite{Schmitz2021}, its application would require a redesign of  available combustion systems.  
Therefore, the use of OMEs in blends with available fossil fuels  is a practical  pathway to reduce pollutant emission retrofitting available combustion systems. 
OMEs show  good miscibility in fossil diesel~\cite{Lin2019,Omari2019}. 
The cetane number of \ce{OME_2} to OME$_5$~\citep{Lautenschutz2016,Deutsch2017} as well as  the flash, boiling and melting points are comparable to those of diesel fuel~\citep{Omari2019,Zheng2013},
indicating their favorable use as fuel blends in compression ignition applications. 
Recently, reduced and detailed kinetic models have been developed for \ce{OME_n} with n~$=$~1-3~\citep{He2018,Sun2017,Li2020,Lin2019,Zhao2020a}, n~$=$~1-4 \cite{Cai2019}, and  n~$=$~1-6~\cite{Niu2021} and validated against experimental data for ignition delay time ~\cite{He2018,Cai2019,DeRas2022} and laminar flame speed ~\cite{Sun2017}. OME blends have been studied under practical  conditions  in a high-pressure vessel~\cite{Goeb2021} and in engine simulations~\citep{Lv2019,Ren2019}. 
Investigations  on soot formation for pure or blended OMEs in canonical flames have been conducted in~\citep{Ferraro2021,Schmitz2021,Tan2021}.
The application of oxymethylene ethers in self-ignition engines in combination with their emission propensity has been studied in several works~\citep{Ren2019, Parravicini2020, Pelerin2020,Pellegrini2013, Barro2018, Huang2018, Leblanc2020, Liu2019} showing that soot emissions are drastically reduced with \ce{OME_n} in combination with conventional diesel fuel while breaking the tradeoff between soot and~\ce{NO_x} emission. 
In our recent work \citep{Ferraro2021}, 
20~\% of \ce{OME_3} blended with ethylene in premixed flames has been found to significantly affect the total number and the size of the particles produced. Specifically, the number of small particles with diameter  $d_p < 5$~nm has been observed to remain unchanged or slightly increase, while the total amount of larger particles and aggregates with diameter $d_p > 20$~nm have been drastically reduced. 
Furthermore, soot structure has been investigated finding a slightly higher  aromaticity for the pure ethylene soot. Particles  produced from  \ce{OME3}-doped flames contained larger amounts of oxygen, mainly in the form of C$=$O bonds. 


Previous studies have been focused  on \ce{OME3} pure and in blends 
with traditional fuels 
(see e.g.~\cite{Ferraro2021,Schmitz2021,Ren2019,Sun2017,Lumpp2011}) due to its better low-temperature fluidity and volatility compared with \ce{OME4} or larger compounds. 
\ce{OME_n} with n $>$ 5 exhibit  a too high melting point, i.e., 18$^\circ$~C for \ce{OME5}~\cite{Omari2019}, while \ce{OME1} is too volatile as a diesel additive~\cite{Lumpp2011}. 
However, OME production is achieved by multiple processes, which lead to a  mix of \ce{OME_n} at different level of polymerization~\cite{Emenike2021,Pelerin2020}. 
A commercially available OME mixture, with physico-chemical properties suitable for diesel applications, will therefore contain  not only short-chain \ce{OME_n} compounds, but also up to 30-wt\%  \ce{OME}$_{\geq5}$ \cite{Omari2019}. 
Hence, understanding the behaviour of the different \ce{OME_n} compounds is of high practical relevance. 

This work focuses on the comparative study of three \ce{OME_n} fuels for n $=$ 2, 3 and 4, combining experimental and numerical investigations to  understand the  features and potential differences in terms of particle emissions at different equivalence ratios, including lightly and highly sooting conditions.

Experimental investigations include quantitative measurements of particulate using in situ laser-based techniques, LIF and LII. A modification of the gas-phase kinetic mechanism employed in \cite{Ferraro2021,Sun2017}, is then formulated to include \ce{OME4} kinetics and validated with experimental data on particle formation.



Similar to \cite{Ferraro2021,Schmitz2021}, the numerical investigations of the premixed sooting flames are conducted with the Conditional Quadrature Method of Moments (CQMOM)~\citep{Salenbauch2017, Salenbauch2018}, based on the physico-chemical soot model by D’Anna et al.~\citep{DAnna2010}. In order to get more insights on the evolution of OME fuels in flames, a detailed kinetic analysis to identify the main reaction pathways of \ce{OME_n} oxidation up to the formation of gas-phase precursors has been conducted.


%% file: parts/exp_setup.tex
Premixed flames burning ethylene as pure fuel and blended with 20~\% of \ce{OME2}, \ce{OME3} and \ce{OME4} with equivalence ratios of $\phi =$ 2.01, 2.16, 2.31 and 2.46 are stabilized on a capillary burner. 
\ce{OME_n}, for n $=$ 2, 3 or 4, are added by replacing some of the ethylene (20~\% of the total carbon fed), being ethylene/air flames the reference. 
Equivalence ratio, cold gas velocity and  total carbon flow rate are kept constant while \ce{OME_n} was added. In order to achieve this, nitrogen and oxygen streams are adapted accordingly.
The same approach has been applied in previous works for other alternative fuels~\cite{Ferraro2021, Conturso2017} and  references therein. 
The flame conditions investigated in this work  are reported in Tab.~\ref{Tab:conditions}.


\begin{table*}[h!] \small
\caption{Flame inflow conditions given in mole fractions. \ce{OME_n} indicates the mole fraction of \ce{OME2}, \ce{OME3} and \ce{OME4} in the corresponding flames, respectively.  }
\begin{tabular}{lccccc}
\hline 
$\phi$ & {} & \ce{C2H4}/\ce{O2}/\ce{N2} & \ce{C2H4}/\ce{OME2}/\ce{O2}/\ce{N2} & \ce{C2H4}/\ce{OME3}/\ce{O2}/\ce{N2} &\ce{C2H4}/\ce{OME4}/\ce{O2}/\ce{N2} \\
\hline
2.01   & \makecell[c]{\ce{OME_n} \\ \ce{C2H4} \\ \ce{N2}  \\ \ce{O2}} & \makecell[c]{-\\0.1234\\0.6925\\0.1841} &\makecell[c]{0.0123\\0.0987\\0.7110\\0.1780} & \makecell[c]{0.0099\\0.0987\\0.7147\\0.1767} & \makecell[c]{0.0082\\0.0987\\0.7172\\0.1759}\\
\hline

 2.16   & \makecell[c]{\ce{OME_n} \\ \ce{C2H4} \\ \ce{N2}  \\ \ce{O2}} & \makecell[c]{-\\0.1313\\0.6862\\0.1824} & \makecell[c]{0.0131 \\	0.1051\\0.7055\\0.1763} &\makecell[c]{0.0105\\0.1051\\0.7093\\0.1751} &\makecell[c]{0.0088\\0.1051\\0.7119\\0.1743}\\
 \hline

 2.31   & \makecell[c]{\ce{OME_n} \\ \ce{C2H4} \\ \ce{N2}  \\ \ce{O2}} & \makecell[c]{-\\0.1392\\0.6800\\0.1808} &\makecell[c]{0.0139\\0.1114\\0.7000\\0.1747 } &\makecell[c]{0.0111\\0.1114\\0.7040\\0.1735} &\makecell[c]{0.0093\\0.1114\\0.7066\\0.1727}\\
\hline

2.46   & \makecell[c]{\ce{OME_n} \\ \ce{C2H4} \\ \ce{N2}  \\ \ce{O2}} & \makecell[c]{-\\0.1469\\0.6739\\0.1792} & \makecell[c]{0.0147\\0.1175\\06946\\0.1732} & \makecell[c]{0.0118\\0.1175\\0.6987\\0.1720} &\makecell[c]{0.0098\\0.1175\\0.7015\\0.1712}\\
\hline
\end{tabular}
\label{Tab:conditions}
\end{table*}

Laser-Induced Emission (LIE) measurements in the 200–550 nm range are  applied to detect particles in the flame, using the fourth harmonic of a Nd:YAG laser at 266 nm as the excitation source~\cite{Ferraro2021,Conturso2016}. 
The emitted spectra are collected with an ICCD camera with a gate of 100 ns, allowing to distinguish between the broad Laser-Induced Fluorescence (LIF) signal, ranging between 300 and 450 nm, and the Laser-Induced Incandescence (LII) following a blackbody curve and evaluated at 550 nm.

%% file: parts/num_setup.tex
\subsection{Gas-phase kinetics}

    


In this work the kinetic mechanism employed in \cite{Ferraro2021,Schmitz2021}, constituted by the detailed mechanism from D'Anna and co-workers~\cite{DAnna2010} combined with the \ce{OME_{1-3}} kinetics from Sun et al.~\cite{Sun2017}, is further developed.  
The kinetic mechanism is extended to cover the decomposition and combustion reactions of \ce{OME4}. Reaction pathways for \ce{OME4} in analogy to smaller OMEs are added following the same reaction coefficients, similarly to the approach adopted by Sun et al.~\cite{Sun2017}. 
Similar to other OMEs, both formation of radicals and unimolecular decomposition are considered. 
Small fragments from \ce{OME4} decomposition follow reaction pathways already established in previous mechanisms~\cite{Sun2017,Ferraro2021}.
In addition, according to considerations made in \cite{Ferraro2021}, a revised oxidation of \ce{CH2O} is proposed according to reaction rates found in the literature. The gas-phase kinetic mechanism extended for the \ce{OME4} fuel consists in total of 757 reactions involving 154 species.
It is worth noting that changes in \ce{CH2O} oxidation does not change the general trend found for different \ce{OME_n} compounds.   
Further details are discussed hereon. 
Temperature profiles measured in the experiments for the pure ethylene flames at different equivalence ratios are imposed in the simulations also for \ce{OME_n} doped flames, similarly to the approach applied in~\cite{Ferraro2021}. Keeping constant equivalence ratio, cold gas velocity and total carbon flow rate when \ce{OME_n} are added allow to have a negligible impact on the temperature profile, as verified for other oxygenated fuels~\cite{Conturso2016}. 

\subsection{Soot modeling}
\vspace{10pt}
A detailed physico-chemical soot formation model~\citep{DAnna2010} combined with the Conditional Quadrature Method of Moments (CQMOM) is employed to study the sooting properties of \ce{OME2} to \ce{OME4}. The numerical approach developed in~\cite{Salenbauch2017} has been successfully applied in atmospheric premixed flames in ~\citep{Salenbauch2017, Salenbauch2018,Ferraro2021}. 
The gas phase kinetics account for  species up to pyrene, while  PAH compounds with a molecular weight larger than pyrene are not treated as individual species but considered as lumped species (\textit{large PAHs}), whose evolution is described by the CQMOM.
The soot model distinguishes between three different particle structures based on their state of aggregation~\citep{DAnna2010}: \textit{large PAHs}, \textit{clusters} and \textit{aggregates}. 
\textit{Clusters} are  spherically shaped, solid soot particles formed by the inception steps involving \textit{large PAHs}. \textit{Aggregates} are fractal-shaped particles generated by the aggregation of several \textit{clusters}.  

The evolution of the chemical and physical properties of each particle class is described by the population balance equation (PBE) for the number density function (NDF) $f(\underline{\xi}; \underline{x}, t)$, which depends on the spatial coordinates $\underline{x}$, the time $t$ and the internal property vector $\underline{\xi}$ with
\begin{equation}
\underline{\xi}=\lbrack \xi_{nc}, \xi_{H/C}, \xi_{stat}, \xi_{typ} \rbrack^{T}. \end{equation}
$\underline{\xi}$ contains two continuous properties $\xi_{nc}$, indicating the number of carbon atoms  with $\xi_{nc} \in \lbrack 0, \infty)$, and $\xi_{H/C}$ describing the carbon to hydrogen ratio with $\xi_{H/C} \in \lbrack 0, 1 \rbrack$.
$\xi_{typ}$ and $\xi_{stat}$ are discrete dimensions representing the type of entities  $\xi_{typ} \in A,A = \lbrace $\textit{large PAHs, clusters, aggregates}$\rbrace$, and $\xi_{stat}$ the chemical reactivity with $\xi_{stat} \in B,B = \lbrace$\textit{stable, radical}$\rbrace$. 


The CQMOM is used to solve the quadrivariate NDF. Following~\cite{Salenbauch2017} the quadrivariate NDF is reformulated in six bivariate NDFs and a set of 36 statistical moments are solved to account for the soot particle evolution. Further details on the numerical approach can be found in~\cite{Salenbauch2017,Salenbauch2018}.

The soot processes, which provide the moment source terms, are formulated based on Arrhenius-rate laws and include growth processes such as the H-Abstraction-\ce{C2H2}-Addition (HACA) mechanism, the resonantly stabilized free radical mechanism or surface growth due to chemical processes.  Nucleation steps for different-sized large PAHs are accounted for, resulting in clusters with various chemical properties as well as oxidation and oxidation-induced fragmentation, dehydrogenation and aggregation processes of several clusters, resulting chain-like formed aggregates.

%% file: parts/results.tex
Firstly, experimental and numerical results for gas-phase species and soot formation in ethylene and  ethylene/\ce{OME_n} flames are presented. Then, reaction pathways flux analyses are performed for the richest condition $\phi=2.46$ to characterize the decomposition of the investigated \ce{OME_n}. Additionally, the influence of selected \ce{OME_n} reactions on formaldehyde formation is evaluated with a sensitivity study.  

\subsection{Gas-phase analysis and soot formation in \ce{OME_{2}}, \ce{OME3}, and \ce{OME4} blended flames}
\vspace{10pt}

\begin{figure*}[!htp]
\centering{
	\includegraphics[width=1\linewidth]{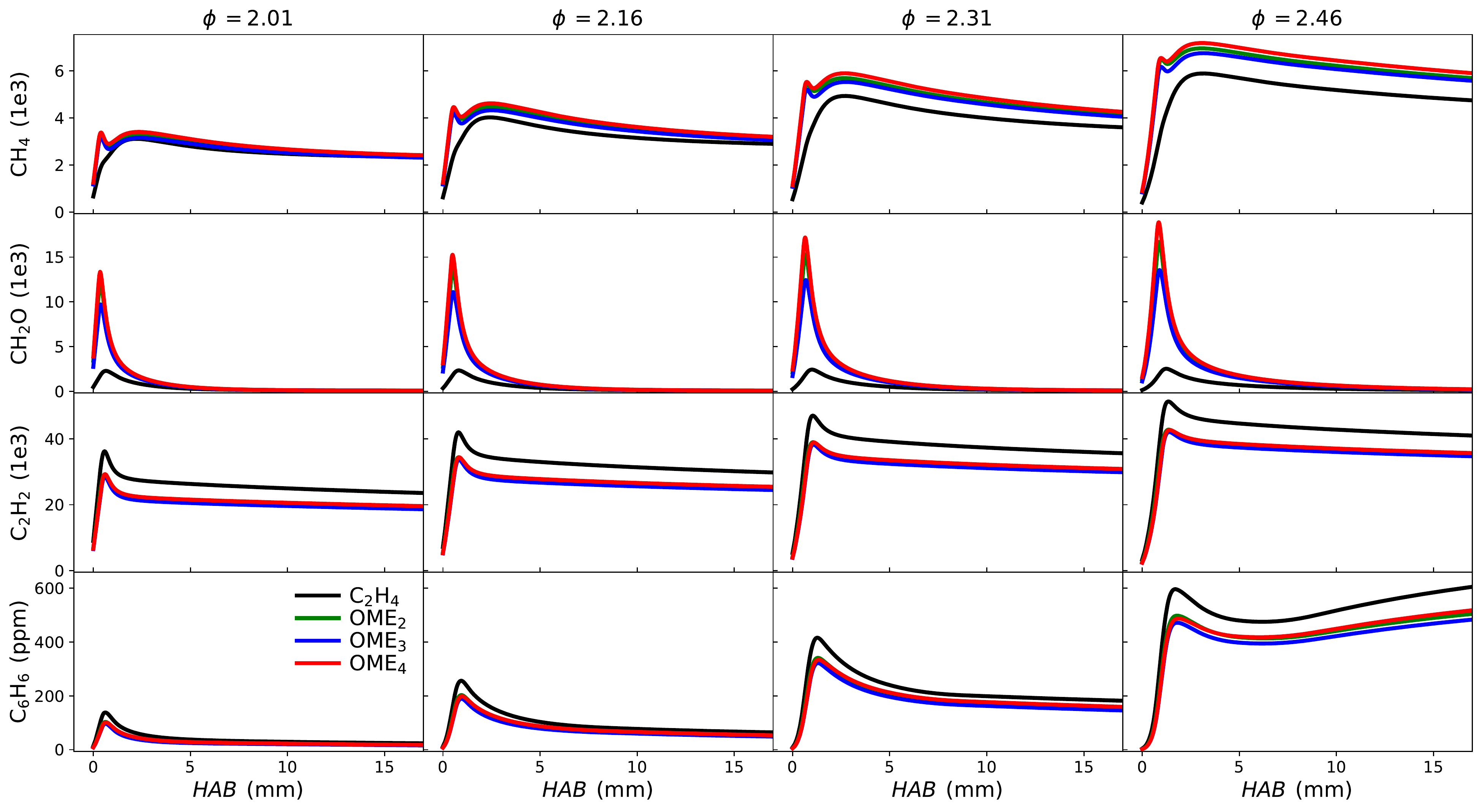}
	\caption{Species mole fraction profiles of \ce{CH4}, \ce{CH2O}, \ce{C2H2} and  \ce{C6H6} as a function of height above the burner HAB obtained for pure ethylene and \ce{OME_n}/ethylene blended mixtures for n $= {2,3,4}$ at different equivalence ratios.}
	\label{fig:species}
	}
\end{figure*}

Modeled mole fraction profiles for selected species in the investigated flames with pure ethylene fuel and \ce{OME_n}/ethylene blended mixtures are here analyzed. Other major species such as \ce{CO}, \ce{H} and \ce{H2}, not shown here for brevity, are almost insensitive to the \ce{OME_n} blending, while a slightly  higher amount of  \ce{CO2} is formed with the addition of \ce{OME_n} compounds as expected. Generally, a better oxidation mechanism for \ce{OME_n} is expected due to the absence of C-C bonds in the fuel structure. Methane (\ce{CH4}) and formaldehyde (\ce{CH2O}), which are key species during the decomposition process of \ce{OME_n,} as well as acetylene (\ce{C2H2}) and benzene (\ce{C6H6}), which are important soot precursor species,  are plotted in Fig.~\ref{fig:species} for the four equivalence ratios  and fuel mixtures against the height above the burner (HAB).    
As expected, \ce{CH4} and \ce{CH2O} increase in the \ce{OME_n} doped flames compared to the pure ethylene flames. In particular, \ce{CH4} is significantly higher in concentration in \ce{OME_n} doped flames as  equivalence ratios increase, suggesting a more effective decomposition process. 
A clear peak in the profiles is visible for \ce{CH2O} close to the inlet, again because of fast oxidation/decomposition pathways, then \ce{CH2O}  quickly decomposes further downstream to smaller species. In general, the preferential pathways leading to the formation of \ce{CH2O} suggest a fast decomposition and a more complete combustion process when \ce{OME_n} are used in comparison to the pure ethylene.

\ce{OME} doping inhibits the formation of species relevant for particle formation and growth such as \ce{C2H2} and \ce{C6H6}. These species lead to the formation of intermediate soot precursors and directly contribute in the soot formation process due to surface growth by HACA process or deposition onto the particle surface (by \ce{C6H6} or larger PAHs). 
\ce{C2H2} is constantly higher in pure ethylene flames across all the equivalence ratios because it is mostly coming from ethylene dehydrogenation pathways. Hence, \ce{C2H2} mainly depends on the amount of ethylene present in the blend disregarding of the equivalence ratio investigated. The effect of \ce{OME_n} is indeed of the first order on this species and a larger effect on particle formation is expected at higher equivalence ratios where surface growth plays a major role. On the other side, the reduction of  \ce{C6H6} increases as equivalence ratio increases, due to relation of \ce{C6H6} formation to intermediate C3 and C4 compounds. The effect of \ce{C6H6} on particle is more complex to follow since it has a direct effect on PAH formation, hence slowing down inception process. This is likely to have an significant effect also at lower equivalence ratio where the inception mechanism is the controlling process for particle formation.  Overall it is interesting to see that both the profile shape and the mole fraction values for all the key species plotted and for all the equivalence ratios show negligible differences for the different \ce{OME_n}.  This suggests similar soot formation behavior of the different \ce{OME_n} compounds. 


\begin{figure*}[!h]
\centering
\includegraphics[width=1\linewidth]{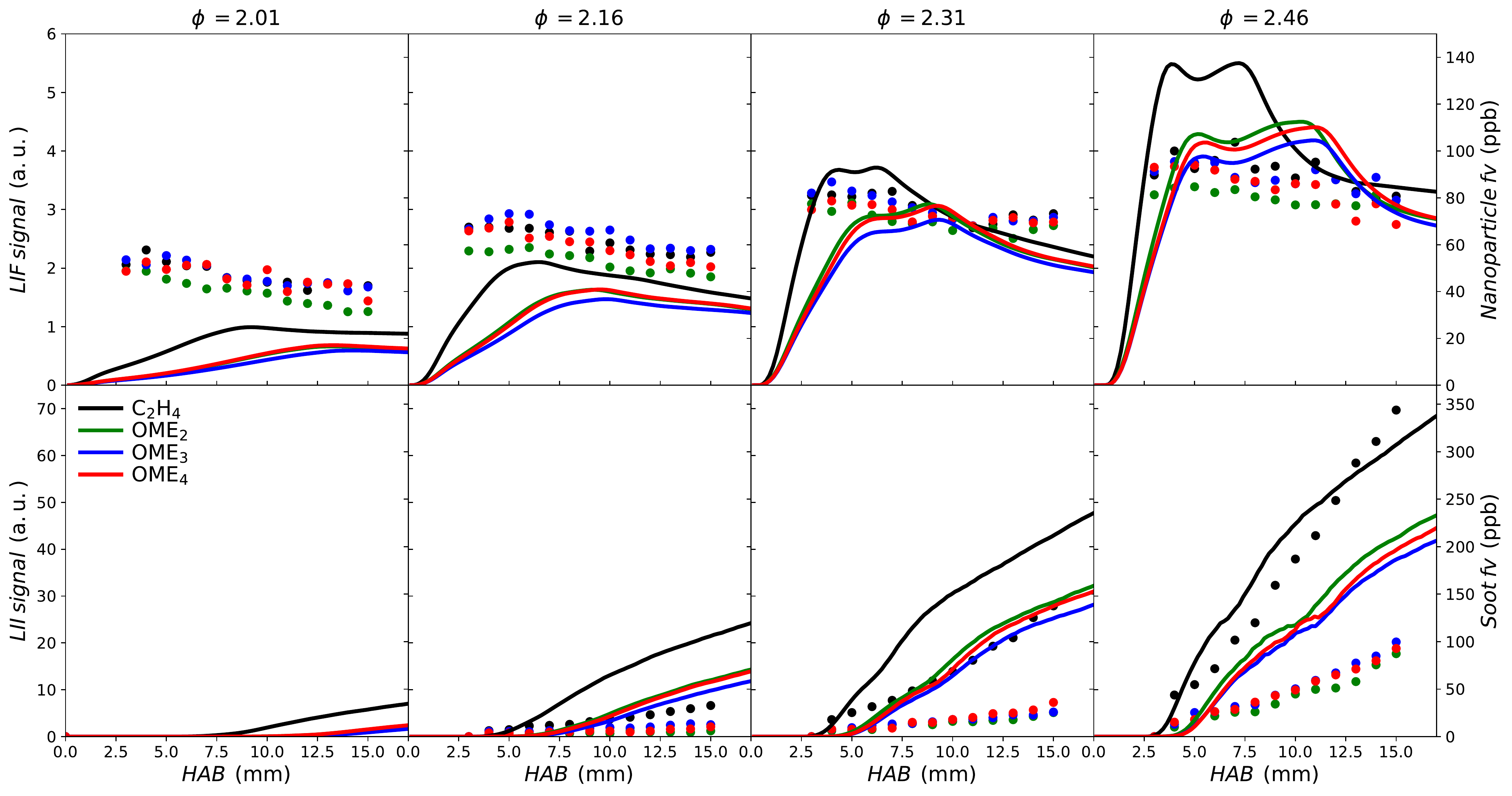}
\caption{Comparison of the simulated nanoparticle volume fraction against the measured LIF signal (top row) and the simulated soot volume fraction against the measured LII signal (bottom row) for the four investigated equivalence ratios and fuel compositions. Lines indicate predicted profiles simulated using the quadvariate CQMOM model whereas dots represent experimental data. The pure ethylene flame is colored black and the \ce{OME_n} doped flames are colored green, blue and red representing \ce{OME_2}, \ce{OME_3} and \ce{OME_4} fuel addition to the flame.}
\label{fig:LIF_LII}
\end{figure*}


Figure~\ref{fig:LIF_LII} shows the modeling results for particles compared with  experimentally measured nanoparticles and soot particles for the investigated equivalence ratios and different \ce{OME_n}-blended mixtures.
As done in previous works, here LIF signal is associated with aromatic hydrocarbons in condensed-phase, i.e. nanoparticles~\cite{Sirignano2017_CombandFlame},
whereas the LII signal is referred to solid soot particles and aggregates. 
In order to better compare modeling results with experimental data, following  previous studies~\cite{Ferraro2021,Salenbauch2018}, the total amount of particles modelled is split depending on particle diameter. 
Specifically,  nanoparticle volume fraction includes particles smaller than the split diameter $d_{p, split}$, while  soot volume fraction includes particles with a diameter larger than $d_{p, split}$. 
Splitting process is here performed  varying the value of split particle diameter, i.e., for $d_{p, split} = 2~\mathrm{nm}$ and $d_{p, split} = 7~\mathrm{nm}$, to reduce the sensitivity of the results on the selected value. 
The plotted volume fraction $fv$ is obtained as the average of both splitting processes. The error bars are not reported here for the sake of clarity, however as shown in previous works the choice of $d_{p, split}$ is not affecting the profile shape nor the absolute value of the nanoparticles volume fraction. 

In previous studies~\cite{Ferraro2021,Schmitz2021} it has been found that \ce{OME3} addition reduces the formation of large soot aggregates significantly, while the formation of smaller nanoparticles is less affected. 
The LIF measurements in Fig.~\ref{fig:LIF_LII} suggest similar trends for \ce{OME2} and \ce{OME4} doped flames. 
There is indeed no noticeable difference of the LIF signal for the pure \ce{C2H4} flame compared to \ce{OME3} and \ce{OME4} doped flames, the \ce{OME2} doped flame exhibits the smallest signal value while still being close to the other results and within the measurement uncertainty. The simulation results of the nanoparticle formation  show minor differences for \ce{OME_n} compounds and therefore reproducing the trends observed in the experiments. The model reproduces a slowing down of the particle formation process. Despite the model slightly overpredicting the nanoparticle peak in the richest ethylene flame, it can reproduce the trend of nanoparticle concentration for ethylene and \ce{OME_n} blends. Overall it is interesting to note that all the OMEs behave similarly to other oxygenated fuels~\cite{Conturso2016}, having a minor impact of the total formation of small nanoparticles, which dominate  particle size distribution in terms of number~\cite{Ferraro2021}.

\begin{figure*}[h]
\centering{
	\includegraphics[width=0.8\linewidth, trim={0 0 0 0},clip]{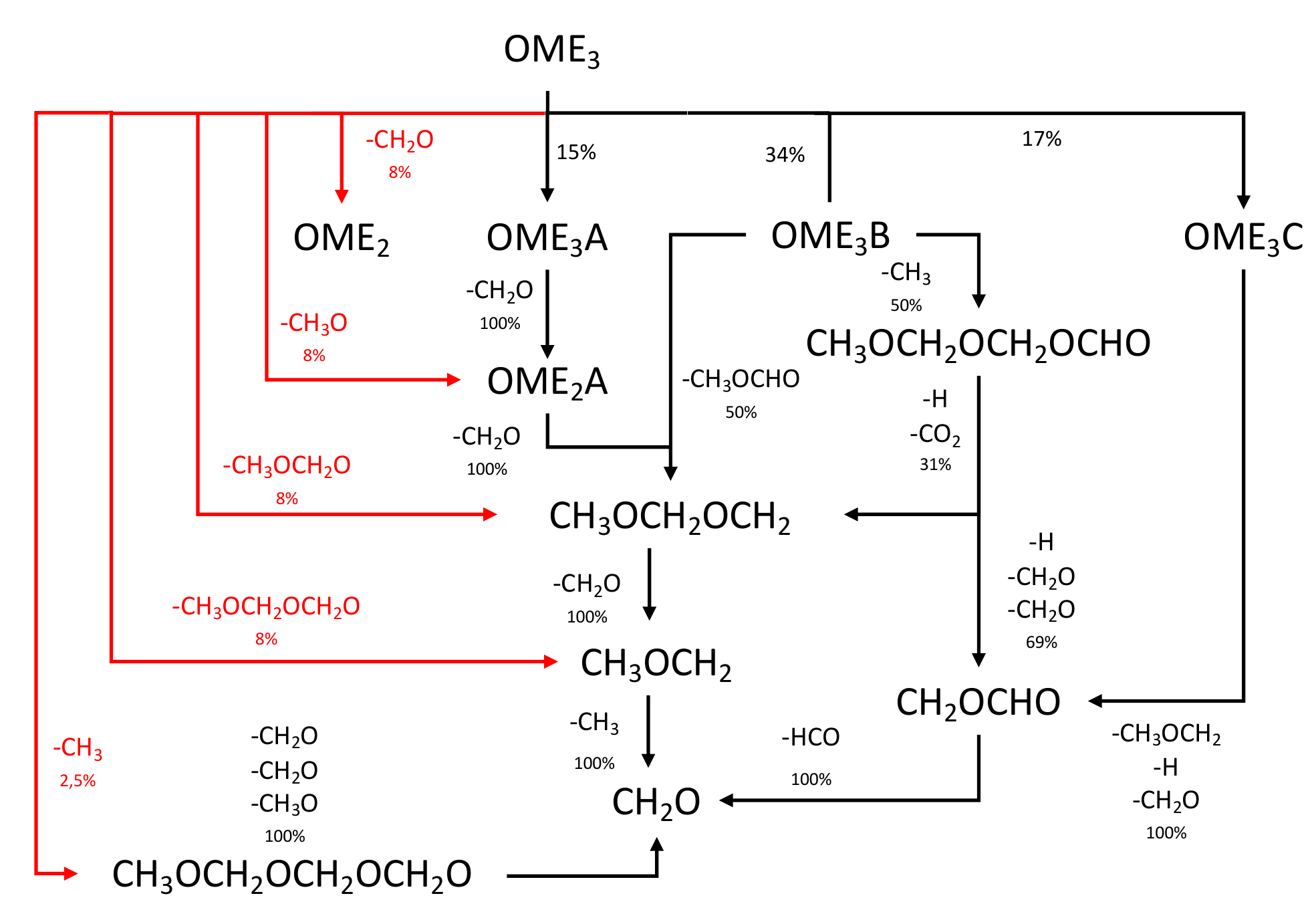}
    \vspace{8 pt}
	\caption{Reaction pathways for \ce{OME3} based on the integrated carbon flux for the \ce{C2H4}/\ce{OME3}/\ce{O2}/\ce{N2} mixture at $\phi=2.46$. The numbers are the flux percent contribution relative to the species on the source side. The red lines indicate the unimolecular decomposition pathways. According to~\cite{Sun2017}, DMM3A: \ce{CH3OCH2OCH2OCH2OCH2}; DMM3B: \ce{CH3OCH2OCH2OCHOCH3}; DMM3C: \ce{CH3OCH2OCHOCH2OCH3}; DMM2A: \ce{CH3OCH2OCH2OCH2}.}
	\label{fig:OME3_path}
	}
\end{figure*}


LII measurements show the onset of soot formation at an equivalence ratio of $\phi = 2.16$ for all \ce{OME_n} mixtures, while no soot  is measured for $\phi = 2.01$. Furthermore,   soot reduction for the three investigated \ce{OME_n} compounds at higher equivalence ratios is   comparable.  
Thus, the experimental results indicate  that the reduction of large  aggregates is similar for mixtures blended with \ce{OME2}, \ce{OME3} and \ce{OME4}.

Regarding the simulation results, the significant reduction of large particles due to \ce{OME2}, \ce{OME3} and \ce{OME4} addition is well captured by the model. Nevertheless, the overall soot reduction predicted by the model is less than the one observed in the experiments. 
Compared to the previous results for \ce{OME3} mixtures~\cite{Ferraro2021}, the formaldehyde  decomposition of the base mechanism has been revised resulting in a larger soot reduction up to 40~\% in terms of large particles in the richest condition for the \ce{OME3} and \ce{OME4} flames, which is closer to the measurement results of around 70~\%, yielding  a significant improvement compared to the  kinetic mechanism used in ~\cite{Ferraro2021}, where soot reduction of approx. 25~\% was numerically achieved. Furthermore, these results confirm the validity of the approach applied to extend the gas-phase kinetics originally developed by Sun et al.~\cite{Sun2017} for n $=$ 1-3 to n $=$ 1-4 for the investigated conditions.
 

\subsection{Reaction analysis}
\vspace{10pt}

\begin{figure*}[h!]
\centering{
	\includegraphics[width=0.8\linewidth]{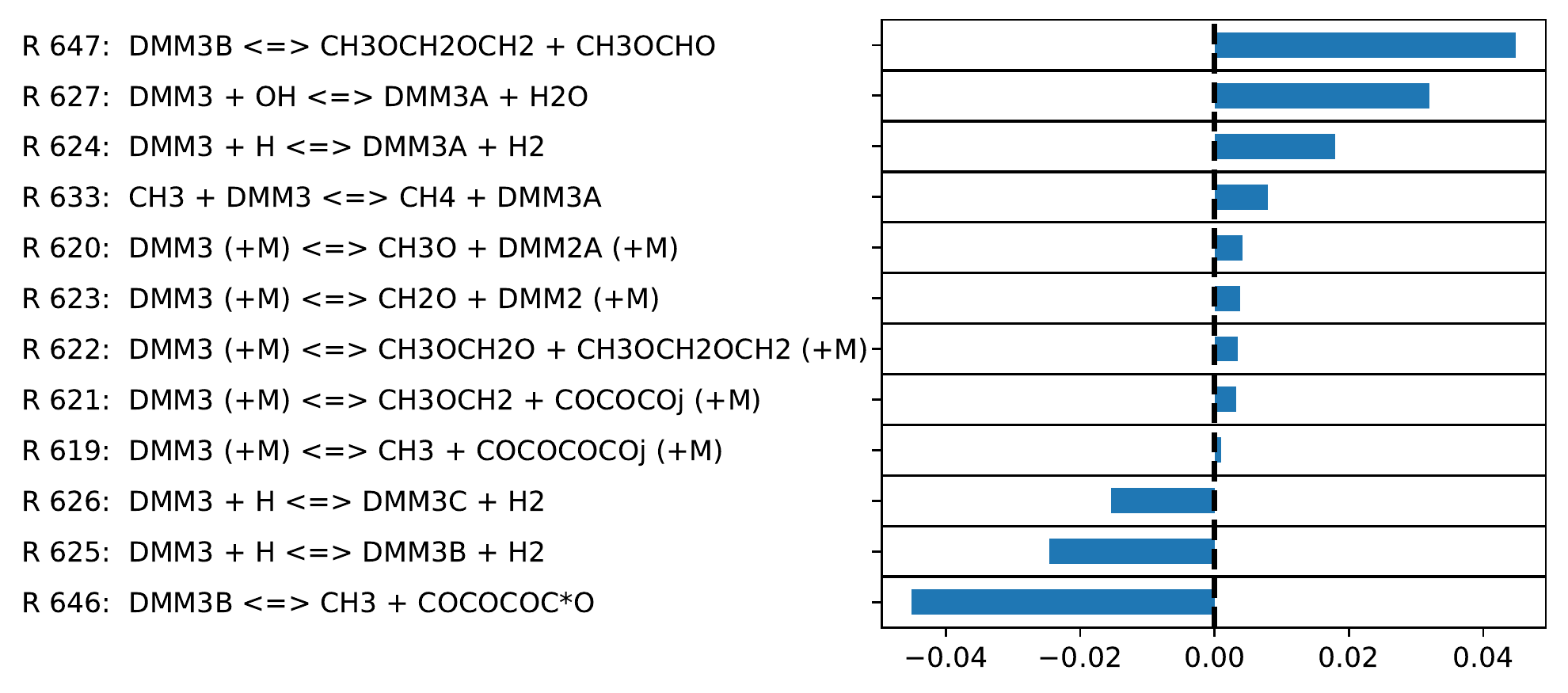}
	\caption{Sensitivity of the \ce{CH2O} mole fraction peak for small perturbations of different reaction rates in the richest \ce{OME3} flame. According to~\cite{Sun2017}, DMM3: \ce{CH3OCH2OCH2OCH2OCH3}; DMM3A: \ce{CH3OCH2OCH2OCH2OCH2}; 
	DMM3B: \ce{CH3OCH2OCH2OCHOCH3}; DMM3C: \ce{CH3OCH2OCHOCH2OCH3}; 
	DMM2: \ce{CH3OCH2OCH2OCH3}; DMM2A: \ce{CH3OCH2OCH2OCH2}; 
	\ce{COCOCOCOj}: \ce{CH3OCH2OCH2OCH2O}; \ce{COCOCOj}: \ce{CH3OCH2OCH2O};
	COCOCOC*O: \ce{CH3OCH2OCH2OCHO}.}
	\label{fig:OME3_sensitivity_v2_CH2O}
	}
\end{figure*}

The gas-phase kinetics is further analyzed to identify the main reaction pathways under rich flame conditions and better understand the reaction pathways for different \ce{OME_n}. 
Figure~\ref{fig:OME3_path} schematically displays the reaction pathways of the \ce{OME3} decomposition to smaller species for a \ce{C2H4}/\ce{OME3}/\ce{O2}/\ce{N2} mixture at an equivalence ratio of $\phi = 2.46$ and temperature $T =$ 1500 K.  The composition of the gas phase is according to Tab.~\ref{Tab:conditions}. The carbon flux indicates that for \ce{OME3} oxidation/decomposition pathways lead rapidly to \ce{CH2O} formation. The importance of this latter in \ce{OME_n} rich combustion is hence crucial and its oxidation/decomposition determines the overall particle reduction. 
Furthermore, it is observed that in the conditions analyzed in this study the unimolecular pathways, almost irrelevant ($<$1~\%) in the conditions analyzed by Sun et al. (cfr. Fig. 5 in \cite{Sun2017}), are here significant. 
Almost 34~\% of the carbon contained in the \ce{OME3} fuel is indeed decomposed through unimolecular decomposition reactions. 
Similar trends have been observed in the reaction pathway analysis for \ce{OME2} ($\approx$40~\%) and \ce{OME4} ($\approx$38~\%) which are not shown here for brevity. 
Large amounts of \ce{CH2O} and other aldehydic compounds are formed during \ce{OME_n} combustion, whereas almost no other oxygenated species are produced. This could be linked with the presence of a specific oxygen functionality onto particles found experimentally  \cite{Ferraro2021}. Further investigation on this point is needed.

Finally, Fig.~\ref{fig:OME3_sensitivity_v2_CH2O} shows the sensitivity of the \ce{CH2O} mole fraction peak in the \ce{OME3} flame with an equivalence ratio of $\phi = 2.46$ for a small change ($k = 0.5$~\%) of the reaction rate coefficients of selected reactions decomposing \ce{OME3} into radicals and smaller species.  
Sensitivity analysis confirms the general picture seen in reaction pathway analysis for \ce{CH2O} formation. Formation of DMM3A radical  and the breaking of \ce{OME3} into small fragments strongly favor \ce{CH2O} formation. On the other side, DMM3B radical and even more DMM3C radical have a negative impact on \ce{CH2O} formation, slowing down the process. DMM3B eventually decomposing towards small fragments (R647) would recover its capability of fast producing \ce{CH2O}. Similar results are obtained for \ce{OME2} and \ce{OME4}. 
Sensitivity analysis for \ce{OME2} shows a strong preference of DMM2A radical for \ce{CH2O} formation without possibility of recovering for DMM2B. On the other side for \ce{OME4}, the only radical that inhibits \ce{CH2O} formation is DMM4C, being DMM4B fast breaking into small fragments and favor \ce{CH2O} formation.


%% file: parts/conclusions.tex
The sooting propensity of different \ce{OME_n} compounds for n $={2,3,4}$ was investigated by combining experimental measurements with numerical simulations. Ethylene flames with equivalence ratios  $\phi = {2.01, 2.16, 2.31, 2.46}$ blended with  20~\% of different \ce{OME_n} compounds were studied.
Experimental LIF and LII measurements for these flames were performed which exhibited similar trends for the soot formation properties of the three \ce{OME_n} investigated. 
The addition of all \ce{OME_n} compounds primarily  reduces the formation of larger soot particles compared to pure ethylene flames, while the formation of smaller particles is less affected. A comparable total amount of soot particle reduction for the three \ce{OME_n} was observed. 
Furthermore, the  kinetic mechanism from Sun et al.~\cite{Sun2017} for  \ce{OME}$_{1-3}$ was extended with \ce{OME4} kinetics. 
Numerical simulations were performed using the detailed mechanism kinetic mechanism  in combination with the CQMOM soot model. The experimentally observed trends of the soot reduction are very well reproduced by the model for all \ce{OME_n} compounds.  Reaction pathway analyses and sensitivity studies for the \ce{OME_n} show the importance of the fuel decomposition under the investigated conditions for \ce{CH2O} formation. Overall, within the investigated conditions, different \ce{OME_n} have shown similar kinetic behavior, suggesting that some differences could arise if the first steps of oxidation/decomposition might become slow enough to be the controlling steps. 





%% file: OME234_LIF_LII_mech.bbl
\begin{thebibliography}{36}
\providecommand{\natexlab}[1]{#1}
\providecommand{\url}[1]{\texttt{#1}}
\expandafter\ifx\csname urlstyle\endcsname\relax
  \providecommand{\doi}[1]{doi: #1}\else
  \providecommand{\doi}{doi: \begingroup \urlstyle{rm}\Url}\fi

\bibitem[Fenard and Vanhove(2021)]{Fenard2021}
Yann Fenard and Guillaume Vanhove.
\newblock {A Mini-Review on the Advances in the Kinetic Understanding of the
  Combustion of Linear and Cyclic Oxymethylene Ethers}.
\newblock \emph{Energy and Fuels}, 35\penalty0 (18):\penalty0 14325--14342,
  2021.

\bibitem[Wang et~al.(2016)Wang, Liu, Ma, Wang, Shuai, and Reitz]{Wang2016}
Zhi Wang, Haoye Liu, Xiao Ma, Jianxin Wang, Shijin Shuai, and Rolf~D. Reitz.
\newblock {Homogeneous charge compression ignition (HCCI) combustion of
  polyoxymethylene dimethyl ethers (PODE)}.
\newblock \emph{Fuel}, 183:\penalty0 206--213, 2016.

\bibitem[Liu et~al.(2016)Liu, Wang, Wang, and He]{Liu2016}
Haoye Liu, Zhi Wang, Jianxin Wang, and Xin He.
\newblock {Improvement of emission characteristics and thermal efficiency in
  diesel engines by fueling gasoline/diesel/PODEn blends}.
\newblock \emph{Energy}, 97:\penalty0 105--112, 2016.

\bibitem[Ferraro et~al.(2021)Ferraro, Russo, Schmitz, Hasse, and
  Sirignano]{Ferraro2021}
Federica Ferraro, Carmela Russo, Robert Schmitz, Christian Hasse, and Mariano
  Sirignano.
\newblock {Experimental and numerical study on the effect of oxymethylene
  ether-3 (OME3) on soot particle formation}.
\newblock \emph{Fuel}, 286:\penalty0 119353, feb 2021.

\bibitem[Schmitz et~al.(2021)Schmitz, Sirignano, Hasse, and
  Ferraro]{Schmitz2021}
Robert Schmitz, Mariano Sirignano, Christian Hasse, and Federica Ferraro.
\newblock {Numerical Investigation on the Effect of the Oxymethylene Ether-3
  (OME3) Blending Ratio in Premixed Sooting Ethylene Flames}.
\newblock \emph{Front. Mech. Eng.}, 7:\penalty0 1--11, 2021.

\bibitem[Tan et~al.(2021)Tan, Salamanca, Pascazio, Akroyd, and Kraft]{Tan2021}
Yong~Ren Tan, Maurin Salamanca, Laura Pascazio, Jethro Akroyd, and Markus
  Kraft.
\newblock {The effect of poly(oxymethylene) dimethyl ethers (PODE3) on soot
  formation in ethylene/PODE3 laminar coflow diffusion flames}.
\newblock \emph{Fuel}, 283:\penalty0 118769, 2021.

\bibitem[Lin et~al.(2019)Lin, Tay, Zhou, and Yang]{Lin2019}
Qinjie Lin, Kun~Lin Tay, Dezhi Zhou, and Wenming Yang.
\newblock {Development of a compact and robust Polyoxymethylene Dimethyl Ether
  3 reaction mechanism for internal combustion engines}.
\newblock \emph{Energy Convers. Manag.}, 185:\penalty0 35--43, 2019.

\bibitem[Omari et~al.(2019)Omari, Heuser, Pischinger, and
  R{\"{u}}dinger]{Omari2019}
Ahmad Omari, Benedikt Heuser, Stefan Pischinger, and Christoph R{\"{u}}dinger.
\newblock {Potential of long-chain oxymethylene ether and oxymethylene
  ether-diesel blends for ultra-low emission engines}.
\newblock \emph{Appl. Energy}, 239:\penalty0 1242--1249, 2019.

\bibitem[Lautensch{\"{u}}tz et~al.(2016)Lautensch{\"{u}}tz, Oestreich,
  Seidenspinner, Arnold, Dinjus, and Sauer]{Lautenschutz2016}
Ludger Lautensch{\"{u}}tz, Dorian Oestreich, Philipp Seidenspinner, Ulrich
  Arnold, Eckhard Dinjus, and J{\"{o}}rg Sauer.
\newblock {Physico-chemical properties and fuel characteristics of oxymethylene
  dialkyl ethers}.
\newblock \emph{Fuel}, 173:\penalty0 129--137, 2016.

\bibitem[Deutsch et~al.(2017)Deutsch, Oestreich, Lautensch{\"{u}}tz, Haltenort,
  Arnold, and Sauer]{Deutsch2017}
Diana Deutsch, Dorian Oestreich, Ludger Lautensch{\"{u}}tz, Philipp Haltenort,
  Ulrich Arnold, and J{\"{o}}rg Sauer.
\newblock {High Purity Oligomeric Oxymethylene Ethers as Diesel Fuels}.
\newblock \emph{Chemie-Ingenieur-Technik}, 89:\penalty0 486--489, 2017.

\bibitem[Zheng et~al.(2013)Zheng, Tang, Wang, Liao, and Wang]{Zheng2013}
Y.~Zheng, Q.~Tang, T.~Wang, Y.~Liao, and J.~Wang.
\newblock {Synthesis of a green fuel additive over cation resins}.
\newblock \emph{Chem. Eng. Technol.}, 36\penalty0 (11):\penalty0 1951--1956,
  2013.

\bibitem[He et~al.(2018)He, Wang, You, Liu, Wang, Li, and He]{He2018}
Tanjin He, Zhi Wang, Xiaoqing You, Haoye Liu, Yingdi Wang, Xiaoyu Li, and Xin
  He.
\newblock {A chemical kinetic mechanism for the low- and
  intermediate-temperature combustion of Polyoxymethylene Dimethyl Ether 3
  (PODE3)}.
\newblock \emph{Fuel}, 212:\penalty0 223--235, 2018.

\bibitem[Sun et~al.(2017)Sun, Wang, Li, Zhang, Yang, Yang, Li, Westbrook, and
  Law]{Sun2017}
Wenyu Sun, Guoqing Wang, Shuang Li, Ruzheng Zhang, Bin Yang, Jiuzhong Yang,
  Yuyang Li, Charles~K. Westbrook, and Chung~K. Law.
\newblock {Speciation and the laminar burning velocities of poly(oxymethylene)
  dimethyl ether 3 (POMDME3) flames: An experimental and modeling study}.
\newblock \emph{Proc. Combust. Inst.}, 36\penalty0 (1):\penalty0 1269--1278,
  2017.

\bibitem[Li et~al.(2020)Li, Herreros, Tsolakis, and Yang]{Li2020}
Runzhao Li, Jose~Martin Herreros, Athanasios Tsolakis, and Wenzhao Yang.
\newblock {Chemical kinetic study on ignition and flame characteristic of
  polyoxymethylene dimethyl ether 3 (PODE3)}.
\newblock \emph{Fuel}, 279:\penalty0 118423, 2020.

\bibitem[Zhao et~al.(2020)Zhao, Li, Xie, Cheng, and Wang]{Zhao2020a}
Yuwei Zhao, Ning Li, Yijing Xie, Yuemeng Cheng, and Xiaochen Wang.
\newblock {Study on chemical kinetic mechanisms of Polyoxymethylene Dimethyl
  Ethers (PODE n )}.
\newblock \emph{IOP Conf. Ser. Mater. Sci. Eng.}, 768\penalty0 (2):\penalty0
  022056, mar 2020.

\bibitem[Cai et~al.(2019)Cai, Jacobs, Langer, vom Lehn, Heufer, and
  Pitsch]{Cai2019}
Liming Cai, Sascha Jacobs, Raymond Langer, Florian vom Lehn, Karl~Alexander
  Heufer, and Heinz Pitsch.
\newblock {Auto-ignition of oxymethylene ethers (OMEn, n = 2–4) as promising
  synthetic e-fuels from renewable electricity: shock tube experiments and
  automatic mechanism generation}.
\newblock \emph{Fuel}, 264:\penalty0 116711, 2019.

\bibitem[Niu et~al.(2021)Niu, Jia, Chang, Duan, Dong, and Wang]{Niu2021}
Bo~Niu, Ming Jia, Yachao Chang, Huiquan Duan, Xue Dong, and Pengzhi Wang.
\newblock {Construction of reduced oxidation mechanisms of polyoxymethylene
  dimethyl ethers (PODE1–6) with consistent structure using decoupling
  methodology and reaction rate rule}.
\newblock \emph{Combust. Flame}, 232:\penalty0 111534, 2021.

\bibitem[{De Ras} et~al.(2022){De Ras}, Kusenberg, Vanhove, Fenard,
  Eschenbacher, Varghese, Aerssens, {Van de Vijver}, Tran, Thybaut, and {Van
  Geem}]{DeRas2022}
Kevin {De Ras}, Marvin Kusenberg, Guillaume Vanhove, Yann Fenard, Andreas
  Eschenbacher, Robin~J. Varghese, Jeroen Aerssens, Ruben {Van de Vijver},
  Luc-Sy Tran, Joris~W. Thybaut, and Kevin~M. {Van Geem}.
\newblock {A detailed experimental and kinetic modeling study on pyrolysis and
  oxidation of oxymethylene ether-2 (OME-2)}.
\newblock \emph{Combust. Flame}, 238:\penalty0 111914, 2022.

\bibitem[Goeb et~al.(2021)Goeb, Davidovic, Cai, Pancharia, Bode, Jacobs,
  Beeckmann, Willems, Heufer, and Pitsch]{Goeb2021}
Dominik Goeb, Marco Davidovic, Liming Cai, Pankaj Pancharia, Mathis Bode,
  Sascha Jacobs, Joachim Beeckmann, Werner Willems, Karl~Alexander Heufer, and
  Heinz Pitsch.
\newblock {Oxymethylene ether – n-dodecane blend spray combustion:
  Experimental study and large-eddy simulations}.
\newblock \emph{Proc. Combust. Inst.}, 38\penalty0 (2):\penalty0 3417--3425,
  2021.

\bibitem[Lv et~al.(2019)Lv, Chen, Chen, Guo, Chen, and Huang]{Lv2019}
Delin Lv, Yingjie Chen, Yaojuan Chen, Xiaoyu Guo, Hui Chen, and Haozhong Huang.
\newblock {Development of a reduced diesel/PODEn mechanism for diesel engine
  application}.
\newblock \emph{Energy Convers. Manag.}, 199:\penalty0 112070, 2019.

\bibitem[Ren et~al.(2019)Ren, Wang, Li, Liu, and Wang]{Ren2019}
Shuojin Ren, Zhi Wang, Bowen Li, Haoye Liu, and Jianxin Wang.
\newblock {Development of a reduced polyoxymethylene dimethyl ethers (PODEn)
  mechanism for engine applications}.
\newblock \emph{Fuel}, 238:\penalty0 208--224, 2019.

\bibitem[Parravicini et~al.(2020)Parravicini, Barro, and
  Boulouchos]{Parravicini2020}
Matteo Parravicini, Christophe Barro, and Konstantinos Boulouchos.
\newblock {Compensation for the differences in LHV of diesel-OME blends by
  using injector nozzles with different number of holes: Emissions and
  combustion}.
\newblock \emph{Fuel}, 259:\penalty0 116166, 2020.

\bibitem[P{\'{e}}lerin et~al.(2020)P{\'{e}}lerin, Gaukel, H{\"{a}}rtl, Jacob,
  and Wachtmeister]{Pelerin2020}
Dominik P{\'{e}}lerin, Kai Gaukel, Martin H{\"{a}}rtl, Eberhard Jacob, and
  Georg Wachtmeister.
\newblock {Potentials to simplify the engine system using the alternative
  diesel fuels oxymethylene ether OME1 and OME3-6 on a heavy-duty engine}.
\newblock \emph{Fuel}, 259:\penalty0 116231, 2020.

\bibitem[Pellegrini et~al.(2013)Pellegrini, Marchionna, Patrini, and
  Florio]{Pellegrini2013}
Leonardo Pellegrini, Mario Marchionna, Renata Patrini, and Salvatore Florio.
\newblock {Emission performance of neat and blended polyoxymethylene dimethyl
  ethers in an old light-duty diesel car}.
\newblock \emph{SAE Tech. Pap.}, 2, 2013.

\bibitem[Barro et~al.(2018)Barro, Parravicini, Boulouchos, and
  Liati]{Barro2018}
Christophe Barro, Matteo Parravicini, Konstantinos Boulouchos, and Anthi Liati.
\newblock {Neat polyoxymethylene dimethyl ether in a diesel engine; part 2:
  Exhaust emission analysis}.
\newblock \emph{Fuel}, 234:\penalty0 1414--1421, 2018.

\bibitem[Huang et~al.(2018)Huang, Liu, Teng, Pan, Liu, and Wang]{Huang2018}
Haozhong Huang, Qingsheng Liu, Wenwen Teng, Mingzhang Pan, Chang Liu, and
  Qingxin Wang.
\newblock {Improvement of combustion performance and emissions in diesel
  engines by fueling n-butanol/diesel/PODE3–4 mixtures}.
\newblock \emph{Appl. Energy}, 227:\penalty0 38--48, 2018.

\bibitem[LeBlanc et~al.(2020)LeBlanc, Sandhu, Yu, Han, Wang, Tjong, and
  Zheng]{Leblanc2020}
Simon LeBlanc, Navjot Sandhu, Xiao Yu, Xiaoye Han, Meiping Wang, Jimi Tjong,
  and Ming Zheng.
\newblock {An Investigation Into OME3 on a High Compression Ratio Engine}.
\newblock In \emph{ASME 2020 Intern. Combust. Engine Div. Fall Tech. Conf.},
  2020.
\newblock ISBN 978-0-7918-8403-4.

\bibitem[Liu et~al.(2019)Liu, Wang, Li, Zheng, He, and Wang]{Liu2019}
Haoye Liu, Zhi Wang, Yanfei Li, Yanyan Zheng, Tanjin He, and Jianxin Wang.
\newblock {Recent progress in the application in compression ignition engines
  and the synthesis technologies of polyoxymethylene dimethyl ethers}.
\newblock \emph{Appl. Energy}, 233-234:\penalty0 599--611, 2019.

\bibitem[Lumpp et~al.(2011)Lumpp, Rothe, Pastotter, L{\"{a}}mmermann, and
  Jacob]{Lumpp2011}
B.~Lumpp, D.~Rothe, C.~Pastotter, R.~L{\"{a}}mmermann, and Eberhard Jacob.
\newblock {Oxymethylene Ethers As Diesel Fuel}.
\newblock \emph{Mtz Worldw.}, 78:\penalty0 34--38, 2011.

\bibitem[Emenike et~al.(2021)Emenike, Michailos, Hughes, Ingham, and
  Pourkashanian]{Emenike2021}
Oluchi Emenike, Stavros Michailos, Kevin~J. Hughes, Derek Ingham, and Mohamed
  Pourkashanian.
\newblock {Techno-economic and environmental assessment of BECCS in fuel
  generation for FT-fuel, bioSNG and OMEx}.
\newblock \emph{Sustain. Energy Fuels}, 5:\penalty0 3382--3402, 2021.

\bibitem[Salenbauch et~al.(2017)Salenbauch, Sirignano, Marchisio, Pollack,
  D'Anna, and Hasse]{Salenbauch2017}
Steffen Salenbauch, Mariano Sirignano, Daniele~L. Marchisio, Martin Pollack,
  Andrea D'Anna, and Christian Hasse.
\newblock {Detailed particle nucleation modeling in a sooting ethylene flame
  using a Conditional Quadrature Method of Moments (CQMOM)}.
\newblock \emph{Proc. Combust. Inst.}, 36:\penalty0 771--779, 2017.

\bibitem[Salenbauch et~al.(2018)Salenbauch, Sirignano, Pollack, D'Anna, and
  Hasse]{Salenbauch2018}
Steffen Salenbauch, Mariano Sirignano, Martin Pollack, Andrea D'Anna, and
  Christian Hasse.
\newblock {Detailed modeling of soot particle formation and comparison to
  optical diagnostics and size distribution measurements in premixed flames
  using a method of moments}.
\newblock \emph{Fuel}, 222:\penalty0 287--293, jun 2018.

\bibitem[D'Anna et~al.(2010)D'Anna, Sirignano, and Kent]{DAnna2010}
Andrea D'Anna, Mariano Sirignano, and John Kent.
\newblock {A model of particle nucleation in premixed ethylene flames}.
\newblock \emph{Combust. Flame}, 157:\penalty0 2106--2115, 2010.

\bibitem[Conturso et~al.(2017)Conturso, Sirignano, and D'Anna]{Conturso2017}
Marielena Conturso, Mariano Sirignano, and Andrea D'Anna.
\newblock {Effect of 2,5-dimethylfuran doping on particle size distributions
  measured in premixed ethylene/air flames}.
\newblock \emph{Proc. Combust. Inst.}, 36:\penalty0 985--992, 2017.

\bibitem[Conturso et~al.(2016)Conturso, Sirignano, and D'Anna]{Conturso2016}
Marielena Conturso, Mariano Sirignano, and Andrea D'Anna.
\newblock {Effect of furanic biofuels on particles formation in premixed
  ethylene-air flames: An experimental study}.
\newblock \emph{Fuel}, 175:\penalty0 137--145, 2016.

\bibitem[Sirignano et~al.(2017)Sirignano, Bartos, Conturso, Dunn, D'Anna, and
  Masri]{Sirignano2017_CombandFlame}
Mariano Sirignano, Daniel Bartos, Marielena Conturso, Matthew Dunn, Andrea
  D'Anna, and Assaad~R. Masri.
\newblock {Detection of nanostructures and soot in laminar premixed flames}.
\newblock \emph{Combust. Flame}, 176:\penalty0 299--308, 2017.

\end{thebibliography}
